%% file: ms.tex
\documentclass[12pt,preprint]{aastex}

\shorttitle{Extragalactic Radio Sources} \shortauthors{M. Gervasi
et al.}

\begin{document}


\title{The contribution of the Unresolved Extragalactic Radio Sources
       to the Brightness Temperature of the sky}


\author{M. Gervasi\altaffilmark{1,2}, A. Tartari, M.
Zannoni\altaffilmark{1}, G. Boella\altaffilmark{2}, and G. Sironi\altaffilmark{1}}
\affil{Physics Department, University of Milano Bicocca,
Piazza della Scienza 3, 20126 Milano Italy}

\email{mario.zannoni@mib.infn.it}

\altaffiltext{1}{also Italian National Institute for Astrophysics,
INAF, Milano.}

\altaffiltext{2}{also Italian National Institute for Nuclear
Physics, INFN, Milano-Bicocca.}


\begin{abstract}
The contribution of the Unresolved Extragalactic Radio Sources to
the diffuse brightness of the sky was evaluated using the source
number - flux measurements available in literature. We first
optimized the fitting function of the data based on number counts
distribution. We then computed the brightness temperature at
various frequencies from 151 MHz to 8440 MHz and derived its
spectral dependence. As expected the frequency dependence can be
described by a power law with a spectral index $\gamma \simeq
-2.7$, in agreement with the flux emitted by the {\it steep
spectrum} sources. The contribution of {\it flat spectrum} sources
becomes relevant at frequencies above several GHz. Using the data
available in literature we improved our knowledge of the
brightness of the unresolved extragalactic radio sources. The
results obtained have general validity and they can be used to
disentangle the various contributions of the sky brightness and to
evaluate the CMB temperature.

\end{abstract}





\section{Introduction}\label{Introduction}

In the past years measurements of the absolute sky emission at
frequency $\nu \sim 1$ GHz have been carried out to evaluate the
brightness temperature ($T_{cmb}$) of the Cosmic Microwave
Background (CMB). Besides the non-trivial problem of assuring an
accurate absolute calibration of the measured signal, we need to
remember that the sky emission is a superposition of different
contributions. After subtracting the local emissions (mainly due
to atmosphere inside the main beam, ground and radio frequency
interferences in the far side-lobes) the sky brightness
temperature ($T_{sky}$) can be written as:

\begin{equation}
T_{sky}(\nu,\alpha,\delta) = T_{cmb}(\nu) + T_{gal}(\nu,\alpha,
\delta) + T_{UERS}(\nu)
\end{equation}

\noindent where $T_{gal}$ is the emission of our galaxy and
$T_{UERS}(\nu)$ the temperature of the unresolved extragalactic
radio sources (UERS). In the present paper we evaluate the UERS
brightness temperature and its frequency dependence.

This paper follows a series of others describing the measurements
of the sky brightness temperature at frequencies close to 1 GHz
gathered by the TRIS experiment with an angular resolution
$FWHM_{TRIS} \sim 20$ deg (\cite[]{TRIS-I}, \cite[]{TRIS-II},
\cite[]{TRIS-III}). The results obtained in the present paper were
used to disentangle the components of the sky brightness and to
evaluate the CMB temperature at the frequencies $\nu =$ 600, 820
and 2500 MHz (\cite[]{TRIS-II}). The aim of this work is to
provide a new estimate of the integrated contribution of UERS to
the diffuse brightness of the sky. An accurate estimate of
$T_{UERS}(\nu)$ is necessary for the TRIS experiment, but also for
all the experiments aimed at the study of the spectral distortions
in the Rayleigh-Jeans tail of the CMB spectrum. Deviations from
the black-body distribution can be present at low frequency, but
the amplitude of the distortions at frequencies around 1 GHz is
nowadays constrained by past experiments at the level of few tens
of mK \cite[]{Fixen_96}.

Experiments like TRIS \cite[]{TRIS-I} can reach a control of
systematics at the level of $\sim$50 mK, a remarkable improvement
if compared to previous measurements at the same frequencies. On
the other hand, relying on the current knowledge of both amplitude
and spectrum of the UERS signal \cite[]{Longair_66}, we can
estimate that at 600, 820, 1400 and 2500 MHz (where CMB
observations have been carried out in the past) the extra-galactic
contribution is respectively $810\pm 180$ mK, $340\pm 80$ mK,
$79\pm 19$ mK and $16\pm 4$ mK (see for example \cite{Sironi_90}).
Using the current 178 MHz normalization \cite[]{Longair_66}, for
state-of-the-art experiments, this means that the uncertainty
associated with the UERS at the lowest frequencies (which are the
most interesting when looking for CMB spectral distortions), is
potentially higher than instrumental systematics. In this paper we
show that by exploiting all the data available in literature we
can significantly improve the present status of our knowledge
about the UERS contribution, and that TRIS-like experiments are
essentially limited by the current technology. New and updated
estimates of the brightness temperature of UERS will be useful
also for feasibility studies of future experiments in this field.

This paper is organized as follows: Section \ref{Sources}
discusses the general properties of the UERS and the data in
literature. In Section \ref{Fit} we describe the procedure to fit
the available number counts; in Section \ref{Brightness} we
calculate the UERS sky brightness and its frequency dependence.
Finally in Section \ref{Discussion} we discuss the implications of
the results obtained for astrophysics and cosmology.


\section{The extragalactic radio sources}\label{Sources}

\subsection{The population of sources}

The unresolved extragalactic radio sources contribute as a blend
of point sources to the measurements of diffuse emission,
especially with poor angular resolution. Actually UERS are an
inhomogeneous collection of quasars, radio galaxies, and other
objects. These can be both compact and extended sources with
different local radio luminosity function, lifetimes and cosmic
evolution. An extensive discussion can be found in
\cite{Longair_78}.

Usually we can distinguish two populations of radio sources: {\it
steep spectrum} sources if $\alpha > 0.5$, and {\it flat spectrum}
sources if $\alpha < 0.5$, where $\alpha$ is the spectral index of
the source spectrum ($S(\nu)\propto \nu^{-\alpha}$). Compact radio
sources, like quasars, have mostly a flat spectrum ($\alpha \simeq
0$) and are most commonly detected at higher frequencies. On the
other hand, extended sources like radio-galaxies, have a steep
spectrum ($\alpha \simeq 0.7-0.8$) and dominate low frequency
counts (see \cite[]{Peacock_81a} and \cite[]{Longair_78}).

{\it Steep spectrum} sources and {\it flat spectrum} sources
contribute in different ways to the number counts. {\it Flat
spectrum} sources are important only at high frequency ($\nu
\gtrsim 2$ GHz). In the same range, {\it flat spectrum} source
counts seem to be comparable to {\it steep spectrum} source counts
for high fluxes, but at low fluxes {\it steep spectrum} sources
are still dominating, as shown for example by \cite{Condon_84a}
and \cite{Kellermann_87} The total number counts have been
successfully fitted using a luminosity evolution model by
\cite{Peacock_81a} and \cite{Danese_87}.

\subsection{Isotropy}

The large scale isotropy of the extragalactic radio sources has
been studied by \cite{Webster_77}. He analyzed several samples
of sources measured in a cube of 1 Gpc-side, getting an upper
limit $\Delta N / N < 3 \%$ on the fluctuation of the source
number. This limit is set by the finite statistics of the sources
contained in the survey: $N \sim 10^{4}$.

\cite{Franceschini_89} have evaluated the fluctuation of the
source counts assuming a poissonian distribution: at $\nu = 5$ GHz
they found fluctuations of the antenna temperature $\Delta T_A /
T_A < 10^{-4}$ over an angular scale $\theta \sim 5$ deg. This
fluctuation rapidly decreases at larger scales. At lower
frequencies the fluctuations increase, but we have at most $\Delta
T_A / T_A < 10^{-2}$ for $\nu=408$ MHz and $\Delta T_A / T_A <
10^{-4}$ for $\nu=2.5$ GHz at an angular scale $\theta \sim 20$
deg.

Due to these considerations, the
contribution of UERS to the sky brightness at large angular scale
is assumed isotropic in the following discussion.
Moreover the radio sources are supposed
to be randomly distributed and the fluctuation in the number
counts is assumed to be poissonian. A possible anisotropic
contribution of UERS is in most cases negligible and limited to
the region of the super-galactic plane (see \cite[]{Shaver_89}).

\subsection{The data set}

Many source-number vs flux distributions have been produced in the
past years: radio source counts at low frequency have been
published even since the Sixties, while deep surveys at higher
frequencies have been performed recently. Compilation of source
counts can be found in several papers (see for example
\cite[]{Condon_84a}, \cite[]{Franceschini_89},
\cite[]{Toffolatti_98}, \cite[]{Windhorst_93}). Most of the data
we used can be extracted from the references included in these
papers.

\begin{figure}
\epsscale{0.9} \plotone{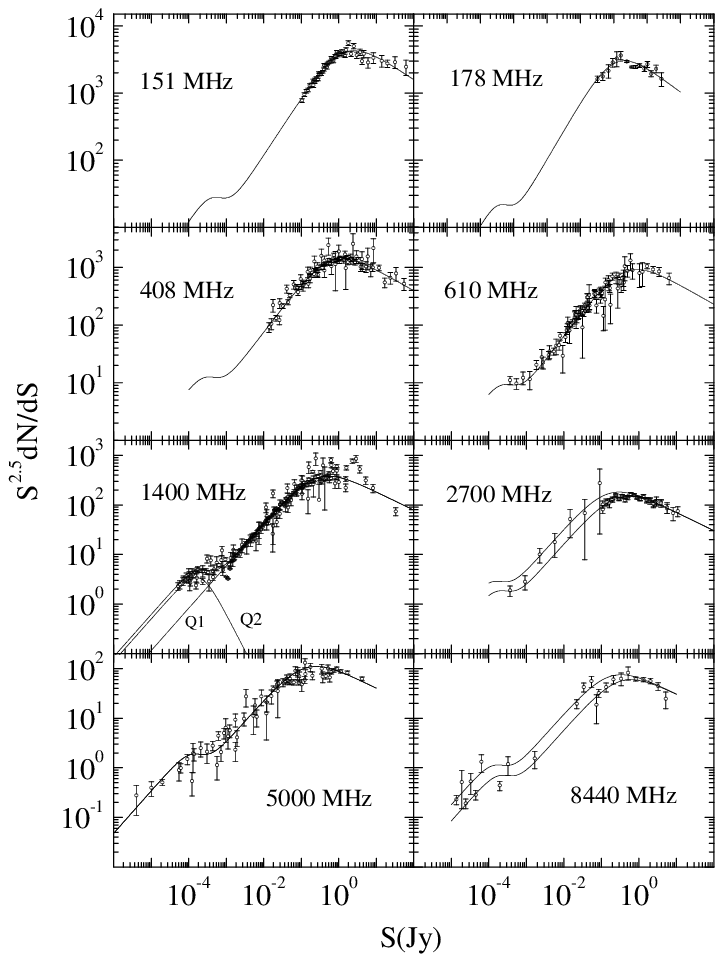} \caption{Data and fit ($Q(S)$) of
source counts at the various frequencies. Data are extracted from
papers listed in Table \ref{tab1}. At $\nu = 1.4$ GHz the two
terms $Q_1$ and $Q_2$ of the fit are also shown. At $\nu = 2.7$
and $\nu = 8.44$ GHz the fit profiles are obtained using the
distribution limits of $A_1$ reported in Table \ref{tab2}.
\label{fig1}}
\end{figure}

We used the counts distributions at the frequencies between 150
and 8000 MHz, spanning a frequency range larger than the one
covered by the TRIS experiment. In literature we found data at
eight frequencies: $\nu=$ 151 MHz, 178 MHz, 408 MHz, 610 MHz, 1.4
GHz, 2.7 GHz, 5.0 GHz, 8.44 GHz. The complete list of the papers
we examined is reported in Table \ref{tab1}. Some of those
measurements were performed at a frequency slightly different from
the nominal one. In this case, the original measurements were
scaled to the nominal frequency by assuming a dependence of the
source flux $S(\nu) \sim \nu^{-0.7}$. Usually the correction is
negligible. The number counts extracted from Table \ref{tab1} is
shown in Figure \ref{fig1}.


\section{Number counts distribution fit}\label{Fit}

The fit was performed on the differential number counts normalized
to the Euclidean distribution of sources $E(S) = S^{5/2}(dN/dS)$,
using the following analytical expression:

\begin{equation}
Q(S) = Q_1(S) + Q_2(S) = \frac{1}{A_1 S^{\varepsilon_1} + B_1
S^{\beta_1}} + \frac{1}{A_2 S^{\varepsilon_2} + B_2 S^{\beta_2}}
\label{Fitequation}
\end{equation}

The best values of the fit parameters are summarized in Tables
\ref{tab2} and \ref{tab3}. This analytical fit function is
empirical. It reproduces the distribution of the experimental data
with some advantages: 1) it is a simple analytical function; 2) it
allows to extend the extrapolation beyond the available data,
because both the amplitude and slope of the tails are well
defined, both at low and high fluxes; 3) it is built as the sum of
two \textit{"populations"} of sources with different emission, but
similar behavior and shape; 4) the fitting procedure is
independent on the source type and evolution but it is applied to
the source counts as currently observed. In addition, this
analytical form is simply a power law, with various indices at
different values of flux, exactly as the observations suggest.

\subsection{High flux distribution fit}

To get the fit we first considered the source distribution at high
fluxes (i.e. we evaluated the parameters of the component
$Q_1(S)$). For this high flux distribution we have data at all the
considered frequencies. The best fit parameters are listed in
Table \ref{tab2}. We found that the values of the spectral indices
$\varepsilon_1$ and $\beta_1$ obtained at all the frequencies are
very similar. We then decided to take a unique weighted average
value for $\varepsilon_1$ and $\beta_1$. The parameter
$\varepsilon_1$ is particularly well constrained by the data at
1400 MHz, where a large set of data with low scatter is available
(see \cite[]{White_97} and Figure \ref{fig1}). Conversely
available data at 178 MHz do not span a flux range wide enough to
constrain the two slopes.

Source counts at 2700 and 8440 MHz are the less accurate among the
full data set we used. At both frequencies there are sets of data
not completely overlapping and the statistics is poor. Therefore,
to take into account these uncertainties, we assumed a uniform
distribution of the parameter $A_1$. At $\nu=2700$ MHz we took
$A_1 = (6 - 12) \times 10^{-4}$, while at $\nu=8440$ MHz we took
$A_1 = (16 - 34) \times 10^{-4}$ (see Table \ref{tab2}). In this
fitting procedure we used all the data collected from the papers
listed in the Table \ref{tab1}. We excluded from the fit only the
seven data points at lowest flux published by \cite{White_97} at
1400 MHz because these data present a roll-off, exactly in the
region where a changing of the slope is expected, but in the
opposite direction. According to  the authors, this roll-off is an
indication of the incompleteness of the survey at the faint-flux
limit.

\subsection{Low flux distribution fit}

\begin{figure}
\begin{center}
\epsscale{0.8} \plotone{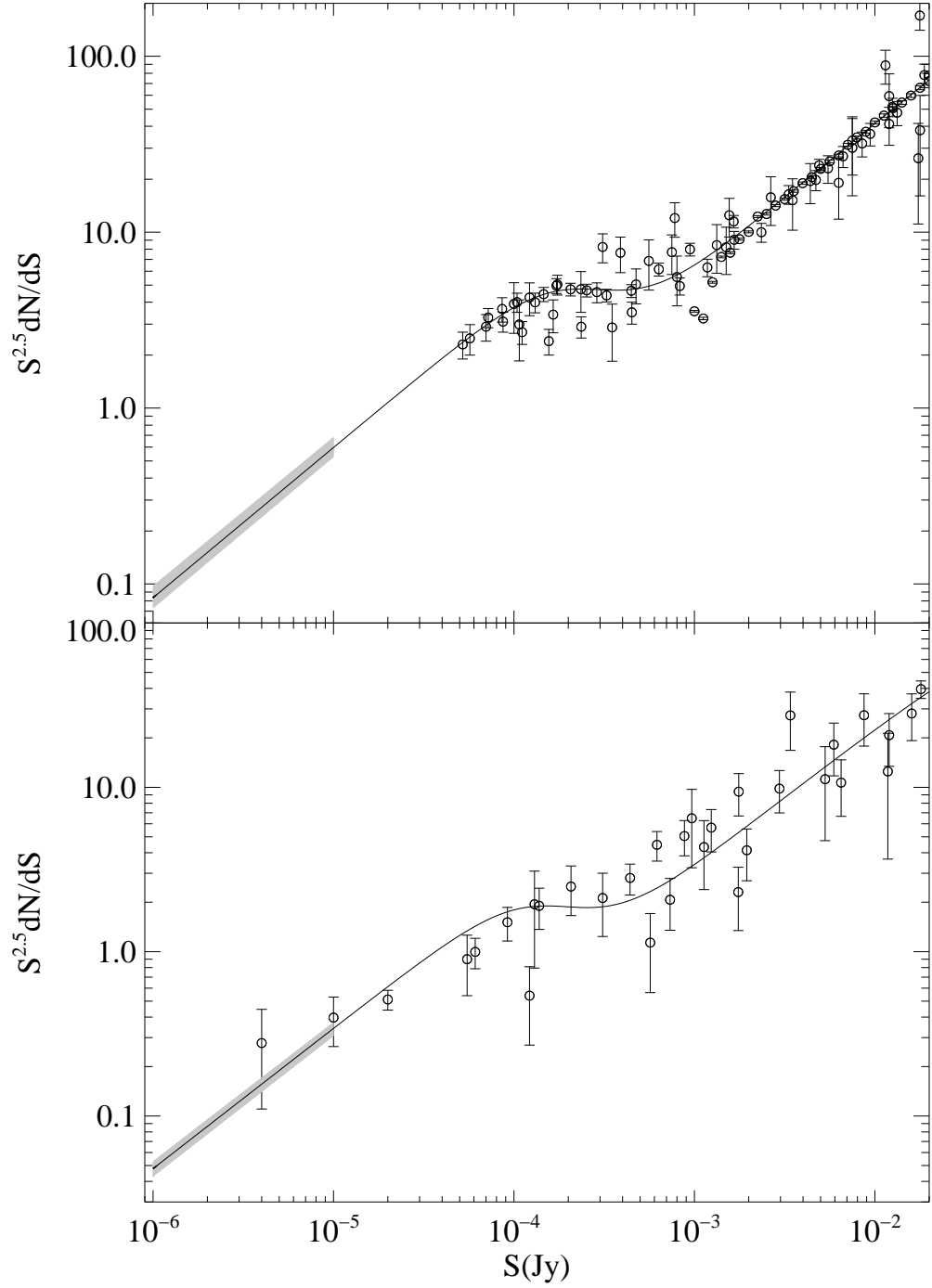} \caption{Data and fit of source
counts at $\nu = 1.4$ and at $\nu = 5$ GHz. The low-flux tail of
the model by \cite{Franceschini_89} is also shown (grey region).
The thickness is an indication of the uncertainty in the
normalization with the experimental data. \label{fig2}}
\end{center}
\end{figure}

We extended the fitting procedure to the counts at low flux, in
order to constrain the distribution $Q_2(S)$. The two parameters
$A_2$ and $\varepsilon_2$ fixed amplitude and slope of the
low-flux tail of the distribution, while $B_2$ and $\beta_2$ are
needed to fit the distribution in the region showing a change in
the slope. Deep counts are available only at 0.61, 1.4, 5 and 8.44
GHz, but at 8.44 GHz the number of experimental points and their
scatter do not allow to fit any parameter. In addition we
considered the model of the low-flux tail published by
\cite{Franceschini_89} at 1.4 and 5 GHz. This source evolution
model fits the data also after the addition of the most recent
measurements both at 1.4 and 5 GHz. The low-flux tail of the model
compared with the experimental data and our fit $Q(S)$ are shown
in Figure \ref{fig2}. This evolution model is able to predict
accurately the slope of the low-flux tail and we used it to get
the value of $\varepsilon_2$, which is independent on the
frequency \cite[]{Franceschini_07}. The values of $\varepsilon_2$,
obtained at 1.4 and 5 GHz, are fully compatible with the average
value of $\varepsilon_1$ previously evaluated. Combining the
estimates at 1.4 and 5 GHz we get $\varepsilon_2 = -0.856 \pm
0.021$.

Since the model by \cite{Franceschini_89} is not able to fix the
amplitude of the number counts, i.e. the parameter $A_2$, we
evaluated this parameter by using the experimental data at 1.4 and
5 GHz. We get: $A_2/A_1 = 0.24 \pm 0.02$ at 1.4 GHz; $A_2/A_1 =
0.30 \pm 0.04$ at 5 GHz. The ratio $A_2/A_1$ is almost independent
on the frequency. The small change is due to the different
contribution of the \textit{flat-spectrum} sources in the total
counts at the two frequencies in the low flux region (below
$10^{-5}$ Jy) and in the high flux region (10 mJy - 1 Jy).

Using the model of \cite{Franceschini_89} and \cite{Toffolatti_98}
at 1.4, 5 and 8.4 GHz, we extrapolated the value of the ratio
$A_2/A_1$ also to the other frequencies. In order to evaluate
$A_2/A_1$ at lower frequencies we estimated the variation of
$A_2/A_1$ at 1.4 GHz excluding from the counts the
\textit{flat-spectrum} sources. This is the extreme condition,
which holds at very low frequencies. In fact the other sources
contributing to the number counts do not change significantly with
frequency. We obtain in this situation that $0.23 \leq A_2/A_1
\leq 0.24$. The same result was obtained starting from data and
model at 5 GHz. Therefore for the frequencies 151, 178, 408 and
610 MHz we take the value obtained at 1.4 GHz, but associating a
larger error bar, due to the uncertainty in the extrapolation
procedure: $A_2/A_1 = 0.24 \pm 0.04$.

We then evaluated the contribution of the {\it flat spectrum}
sources in the total counts at 8.44 GHz (see \cite{Toffolatti_98})
in comparison with the counts at 5 GHz. In this way we estimate at
8.44 GHz the value $A_2/A_1 = 0.31 \pm 0.04$. At 2.7 GHz we took a
value constrained by the results obtained at 1.4 and 5 GHz:
$A_2/A_1 = 0.24 - 0.30$.

We finally estimated the $B_2$ and $\beta_2$. They are important
just to define the shape of the distribution in the region showing
a change in the slope. At 1.4 GHz data are accurate enough to
constrain both parameters, but $\beta_2$ can not be constrained at
the other frequencies. Since the accuracy of these two parameters
is not important for the calculation of the integrated brightness
temperature, we assumed for them the average value for all the
frequencies.

The summary of the best values of all the parameters of $Q_2(S)$
is shown in Table \ref{tab3}. The number counts and the function
which fit them are shown in Figure \ref{fig1}. In conclusion we
can note that: 1) $A_1$ and $B_1$, the two frequency dependent
parameters of the fit, take different values at each frequency.
The same is true for the ratio $A_2/A_1$. 2) The power law indices
($\varepsilon_1$, $\beta_1$, $\varepsilon_2$ and $\beta_2$) are
frequency independent and we take a common value at all the
frequencies.


\section{The UERS contribution to the sky diffuse emission}\label{Brightness}

\subsection{Evaluation of the diffuse emission}

The contribution of the UERS ($B_{UERS}(\nu)$) to the sky
brightness is evaluated by integrating the function $S(dN/dS)$
from the largest flux ($S_{max}$) of the measured sources down to
the lowest fluxes ($S_{min}$) corresponding to the faintest
sources:

\begin{equation}
B_{UERS}(\nu) = \int^{S_{max}}_{S_{min}} \frac{dN}{dS}(\nu) \cdot
S \ dS
\end{equation}

The brightness temperature $T_{UERS}(\nu)$ is by definition:

\begin{equation}
T_{UERS}(\nu) = B_{UERS}(\nu) \frac{\lambda^2}{2 \ k_B},
\end{equation}

\noindent $k_B$ being the Boltzmann constant. The values of
$T_{UERS}$ at the eight frequencies we considered are
summarized in Table \ref{tab4}.

\begin{figure}
\epsscale{1.0} \plotone{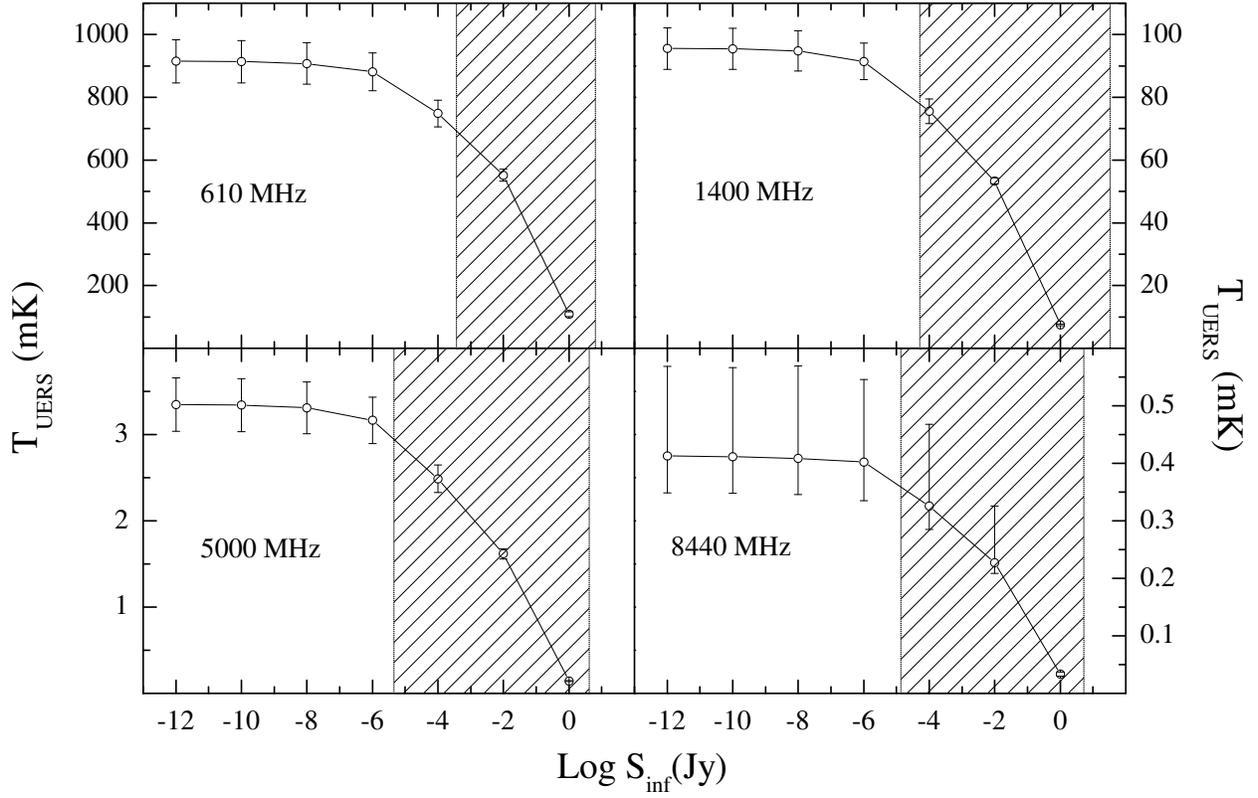} \caption{The integrated brightness
temperature at selected frequencies is shown \textit{vs} the
lowest flux value of source counts ($S_{inf}$). The upper limit of
the integration is fixed at 100 Jy. The region where experimental
data are available is also shown (shadowed region). \label{fig3}}
\end{figure}

From the observations we have $S_{max} \sim 10^2$ Jy (as measured
at 151, 408 and 1400 MHz) and $S_{min} \sim 10^{-6}$ Jy (as
measured in the deepest counts at 5 GHz). While sources at higher
fluxes if present in the surveyed sky region can be easily
measured, the limit at low flux is set by the confusion limit or
by the observation completeness. In other words there is no sharp
limit at low flux in the population of the sources. We extended
the integration down to very faint limits, several orders of
magnitude below the faintest detected sources ($S_{min} \sim
10^{-6}$ Jy). When the integration is extended down to $S_{min}
\sim 10^{-12}$ Jy \ the brightness increase by $3-4$ \% and then
the value converges. This increment is comparable with the total
uncertainty we get on the value of the brightness, as shown in
Figure \ref{fig3}.

We extended the integration also to higher values of the flux, in
order to test also this integration limit. Increasing $S_{max}$
well beyond the flux of the strongest sources observed, the
integral change by less than $0.5\%$ and quickly converges. This
is a consequence of the very low statistics of sources at the
highest fluxes. This is a confirmation that the large scale
brightness is actually not sensitive to the upper limit of
integration.

\subsection{Evaluation of the uncertainty}

The error budget takes into account both the fit uncertainties and
the number counts fluctuations over the observed sky region. The
fit uncertainties were evaluated by means of Monte Carlo
simulations. For each parameter we considered a gaussian
distribution with standard deviation as reported in Tables
\ref{tab2} and \ref{tab3}. Only for the values of $A_1$ at 2.7 and
8.44 GHz and for $A_2/A_1$ at 2.7 GHz we assumed a uniform
distribution inside the interval reported in Tables \ref{tab2} and
\ref{tab3}. The error bars of the parameters of the fit (in
particular $A_2/A_1$ and $\varepsilon_2$) include the uncertainty
on the extrapolation at the lowest fluxes for the various
frequencies. We underline that, as shown in Figure \ref{fig3}, the
contribution to the brightness temperature of the low-flux tail
(below $\sim 10^{-6}$ Jy) is lower than the overall uncertainty
reported in Table \ref{tab4}.

\begin{figure}
\epsscale{1.0} \plotone{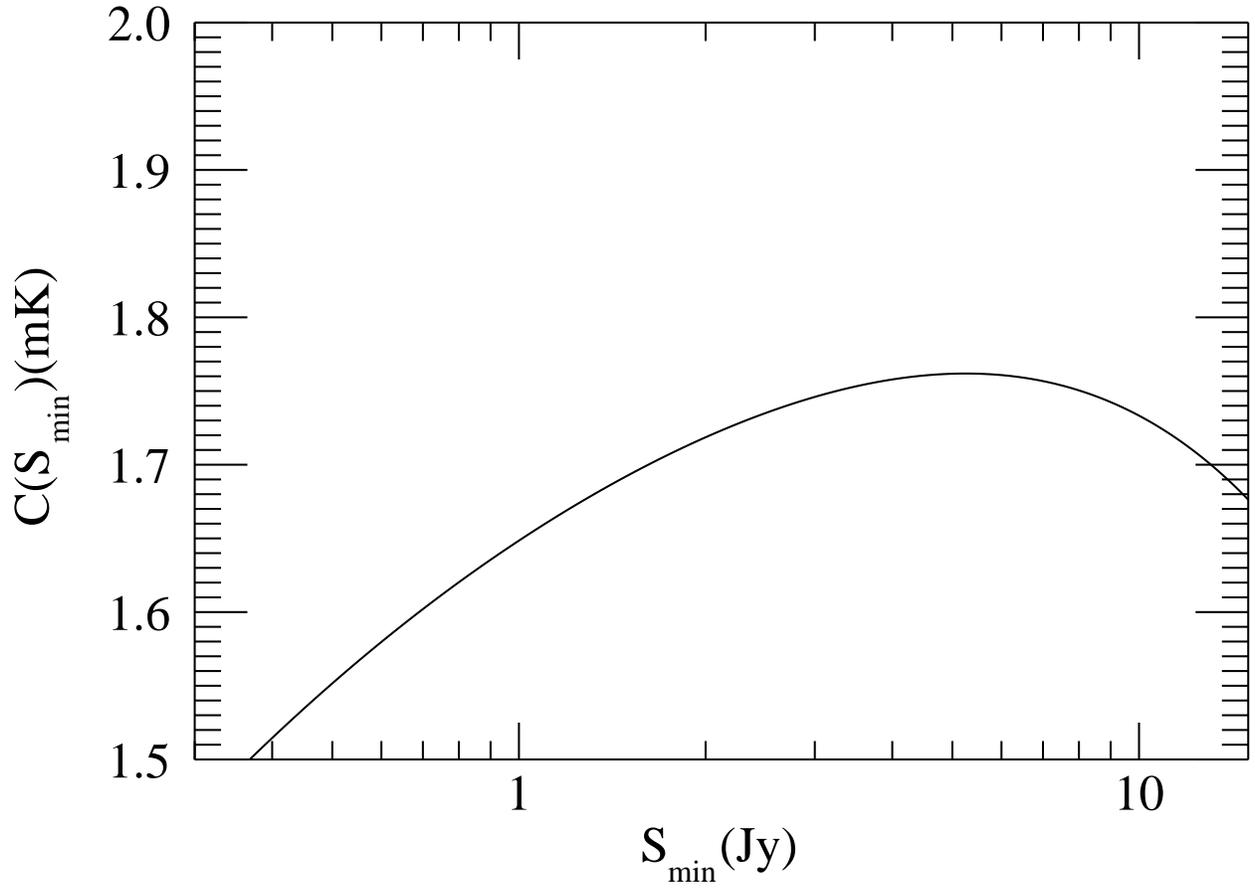} \caption{Frequency: 1400 MHz. The
maximum of the plotted function $C(S_{min})$ \textit{vs} $S_{min}$
sets an upper limit to the contribution of the source number
fluctuation to the brightness temperature uncertainty in a beam
large 0.1 sr (see the text).} \label{fig4}
\end{figure}

The statistical fluctuations of the sources' number counts have no
effect in a large part of the distribution because the number of
sources is quite large. We concentrated on the effect of the
fluctuation of the few sources with the highest flux. We
considered these sources randomly distributed and therefore their
fluctuation is Poissonian. We evaluated the fluctuation of the
brightness in a patch of the sky corresponding to the beam of
TRIS: $\Omega_{TRIS} \sim 0.1$ sr. The upper limit of the
contribution to the temperature uncertainty due to the fluctuation
in the number of sources is directly measured by the maximum of
the function

\begin{equation}\label{}
C(S_{min})=\frac{\lambda^2}{2k_B\Omega_{TRIS}}
\frac{\int_{S_{min}}^{100Jy}\frac{dN}{dS}SdS}
{\int_{S_{min}}^{100Jy}\frac{dN}{dS}dS}
\end{equation}

\noindent plotted in Figure \ref{fig4} for the specific case of
the 1400 MHz data. For all the frequencies this maximum falls
around $\sim 5$ Jy, and its value is from 2 to 6 times smaller
than the corresponding values reported in Table \ref{tab4}.
Therefore for every frequency, and over the full flux range, the
error of the brightness temperature is dominated by the
statistical uncertainties of the fit parameters.

The relative error bar of the brightness temperature is $6-7\%$ at
151, 408, 610 and 1400 MHz increasing up to $9\%$ at 5000 MHz. At
178 MHz the available measurements are few and old and the
obtained error bar is $13\%$. At 2700 and 8440 MHz both the
quantity and the quality of the data do not allow accurate
estimates of the parameters of the fit and we get an uncertainty
of $25-30\%$.

\subsection{Frequency dependence}

\begin{figure}
\epsscale{1.0} \plotone{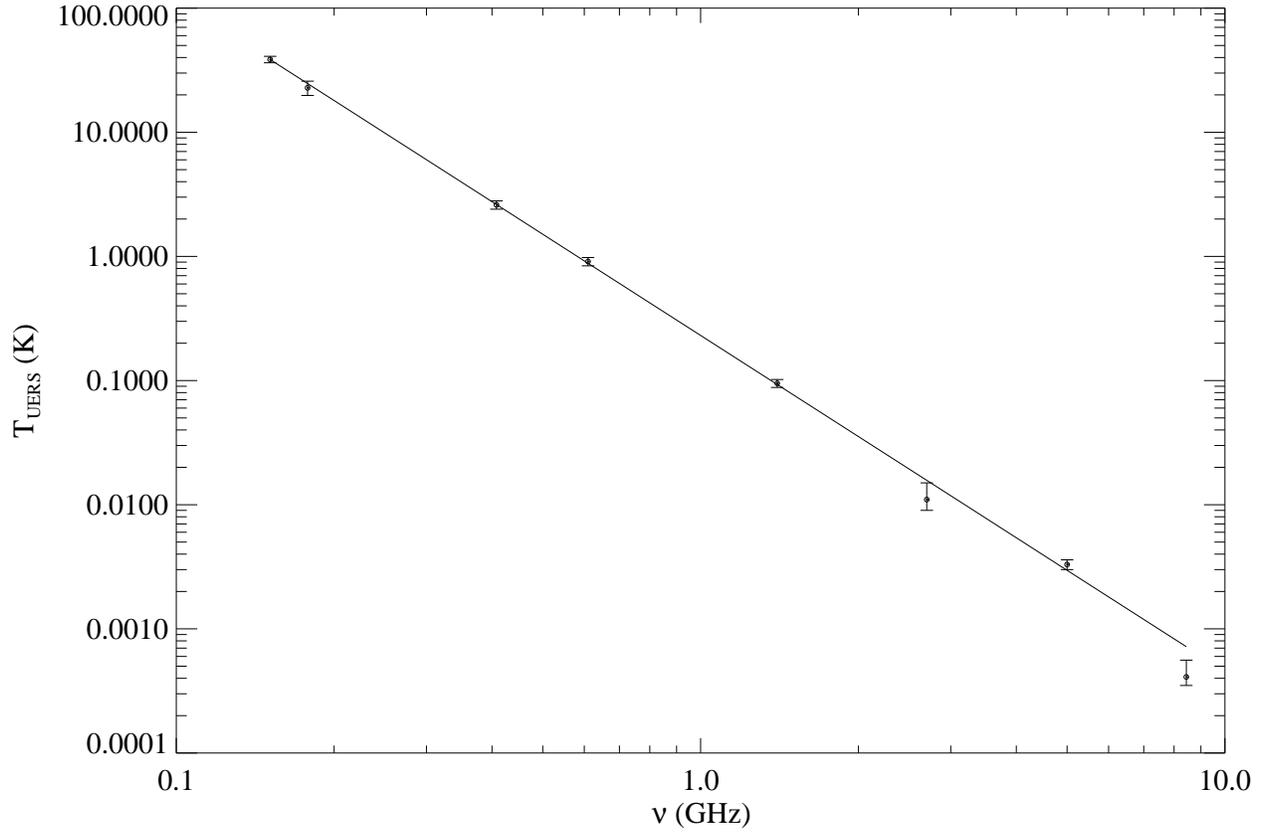} \caption{Frequency spectrum of the
brightness temperature of the unresolved extragalactic radio
sources. The continuous line is \textit{FIT1} (see the text).
\label{fig5}}
\end{figure}

The integrated contribution of the UERS to the brightness of the
sky at the various frequencies is shown in Figure \ref{fig5}. The
distribution can be fitted by a power law:

\begin{equation}
T_{UERS}(\nu) = T_0 \Bigl( \frac{\nu}{\nu_0}\Bigr)^{\gamma_0}
\end{equation}

\noindent Setting $\nu_0 = 610$ MHz (chosen because it is
close to one of the channels of the TRIS experiment), we obtain
the best fit of $T_0$ and $\gamma_0$ shown in Table \ref{tab5}
(\textit{FIT1}). As shown in Figure \ref{fig5}, in spite of the
large error bars at 2700 and 8440 MHz, scatter of the data points is
limited. The fit of a single power law, done excluding the data with
the largest uncertainty (at $\nu=178$, 2700 and 8440 MHz), gives
the values of $T_0$ and $\gamma_0$ shown in Table \ref{tab5}
(\textit{FIT2}). Now the scatter of the experimental data is much
smaller, as shown by the value of the reduced $\chi^2$.

In both cases the results obtained are fully compatible with the
slope of the {\it steep spectrum} sources, which are therefore the
main contributors to the source counts. However it is interesting
to check the contribution of the {\it flat spectrum} sources
especially at high frequencies. We assumed a {\it flat spectrum}
component with fixed slope $\gamma_1 = -2.00$ and a  dominant {\it
steep spectrum} component with slope $\gamma_0 = -2.70$:

\begin{equation}
T_{UERS}(\nu) = T_0 \Bigl( \frac{\nu}{\nu_0}\Bigr)^{\gamma_0} +
T_1 \Bigl( \frac{\nu}{\nu_0}\Bigr)^{\gamma_1}
\end{equation}

\begin{figure}
\epsscale{1.0} \plotone{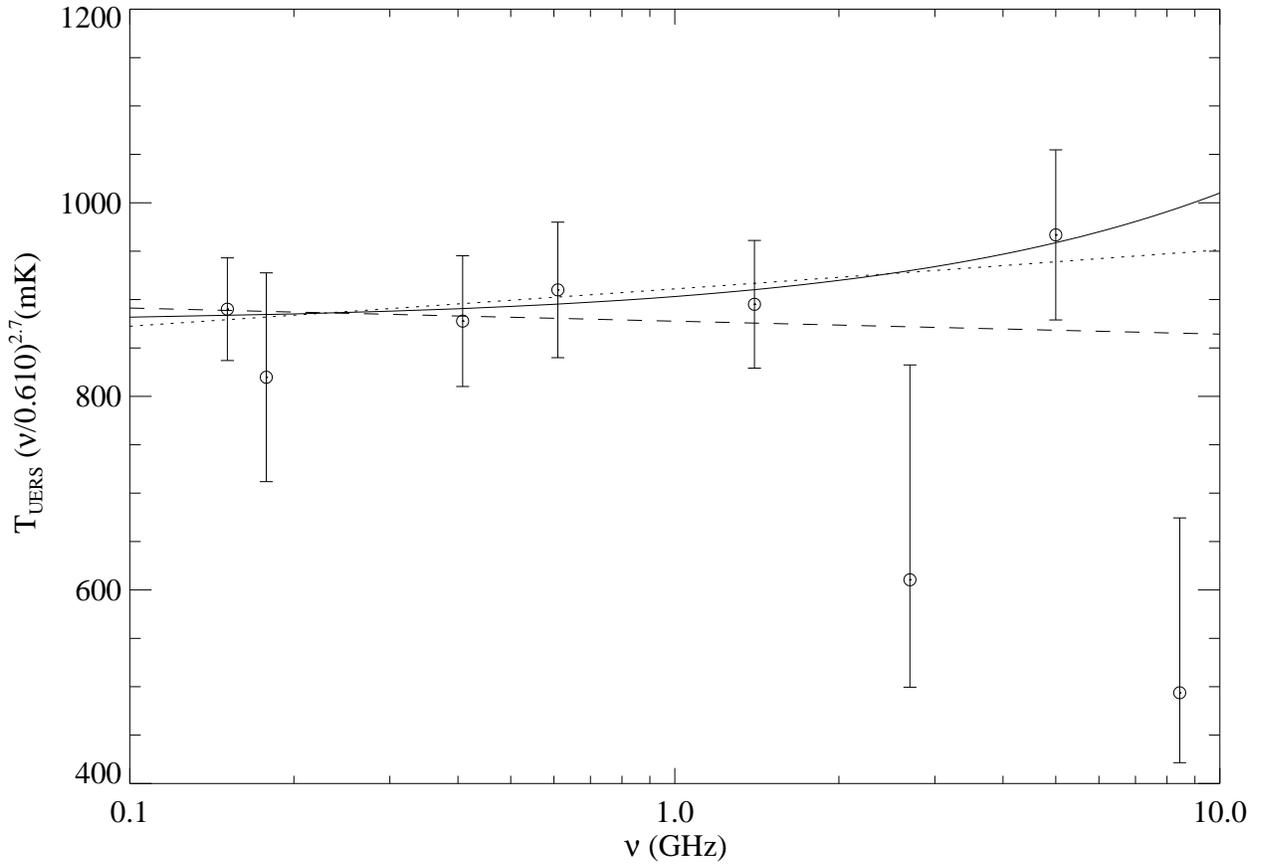} \caption{Brightness temperature of
the extragalactic radio sources. The values are multiplied by the
power law $(\nu/0.61GHz)^{2.70}$. The fit curves discussed in the
text are also shown: \textit{FIT1} (dashes); \textit{FIT2} (dots);
\textit{FIT3} (continuous line). \label{fig6}}
\end{figure}

\noindent The results of the fit of $T_0$ and $T_1$ are shown in Table
\ref{tab5} (\textit{FIT3}). Doing so we can get an estimate of
the contribution of the {\it flat-spectrum} component in the
number counts at the various frequencies. The value of
$T_{UERS}(\nu) (\nu /\nu_0)^{2.70}$ \ at the frequencies analyzed
is shown in Figure \ref{fig6}, together with the power law best
fits. Figure \ref{fig6} shows that the two last fits
(\textit{FIT2} and \textit{FIT3}) are equivalent within the error
bars.

\section{Discussion}\label{Discussion}

\subsection{Analytical form of the fit function}

As discussed in Section \ref{Fit}, the fit function $Q(S)$ is a
power law distribution taking different slopes and amplitudes in
three different regions of the sources' flux. We prefer to deal
with a power law (like $Q(S)$) instead of a polynomial fit, as
performed by \cite{Katgert_88} and \cite{Hopkins_03}. A polynomial
fit can be better adapted to the experimental points, but its
validity range is restricted to the region covered by experimental
points. Therefore it is not possible to use a polynomial fit to
extrapolate the distribution outside this range. Conversely our
fit function can be extrapolated because amplitude and slope of
the tails are well defined and take into account the shapes
expected from the counts evolution models
\cite[]{Franceschini_89}.

According to \cite{Longair_78} we should expect a broadened
maximum in the distribution of the source counts with increasing
frequency. This indication seems to be confirmed looking at the
differential counts shown in (\cite[]{Kellermann_87} and
\cite[]{Condon_84a}). In spite of these expectations we have
obtained a good fit of the counts at the various frequencies,
using the same function with the same slopes, as shown Figure
\ref{fig1}. Part of this effect is probably hidden by the larger
scatter of the high frequency differential counts. In any case the
broadening of the maximum could be marginally observed above 1400
MHz and the effect is not relevant for the calculation of the
contribution to the sky brightness.

\subsection{Frequency dependence}

The spectral dependence of the brightness temperature $T_{UERS}$
follows the expectations at low frequency where the number counts
are dominated by the sources with steep spectrum: $\alpha \sim
0.7$. At high frequency the situation is more complex. We could
expect a flattening, because at these frequencies the number
counts of {\it flat-spectrum} sources begin to be important. It
will probably appear at frequencies higher than 1400 MHz.
Unfortunately the available data are not accurate enough to
constrain the fit parameters (in particular $A_1$) at 2700 and
8440 MHz. The values obtained at these two frequencies, lower than
the values expected looking at the other frequencies, could also
be an indication of incompleteness of the surveys (see Figure
\ref{fig6}). Conversely at 5000 MHz data are better and numerous.
We obtain a value for $T_{UERS}$ fully consistent with the slope
of the data at low frequency, where {\it steep-spectrum} sources
are dominating, even if there is a marginal indication of a
spectral flattening (see Figure \ref{fig6}). In fact when we fit
the data adding the contribution of the {\it flat-spectrum}
sources we get an estimate of this contribution which is $T_1/T_0
\simeq 2\%$ at $\nu=610$ MHz and $T_1/T_0 \simeq 9\%$ at $\nu=5$
GHz (see \textit{FIT3} in the Table \ref{tab5}).

\subsection{Previous estimates of $T_{UERS}$}

The values of brightness temperature shown in Table \ref{tab4}
have error bars of roughly 7\% at most frequencies. The
uncertainty is a bit larger at 178 MHz and even worse at 2700 and
8440 MHz, because of the quality of the number counts data. So far
very few estimates of $T_{UERS}$ have been published (see
\cite[]{Longair_66}, \cite[]{Wall_90} and \cite[]{Burigana_04}),
and the frequencies covered were limited and sometimes the
uncertainty was not quoted. Our results are in agreement with the
values previously estimated by \cite{Longair_66}, $T_{UERS}(178 \
MHz)=23\pm5$ K, and by \cite{Wall_90}, $T_{UERS}(408 \
MHz)\simeq2.6$ K, $T_{UERS}(1.4 \ GHz)\simeq0.09$ K, and
$T_{UERS}(2.5 \ GHz)\simeq0.02$ K, but our error bars are
definitely smaller. The accuracy of the estimated values of
$T_{UERS}$ is particularly important if this contribution is to be
subtracted to calculate the value of the CMB temperature at low
frequency. Table \ref{tab4} suggests that the value of $T_{UERS}$
needs to be accurately evaluated up to a frequency of several GHz,
because its value is not negligible.

\section{Conclusions}

We used the source number - flux measurements in literature to
evaluate the contribution of the Unresolved Extragalactic Radio
Sources to the diffuse brightness of the sky. We analyzed the
count distributions at eight frequencies between 150 and 8000 MHz,
spanning over the frequency range partially covered by the TRIS
experiment (see \cite[]{TRIS-I}): $\nu=$ 151 MHz, 178 MHz, 408
MHz, 610 MHz, 1.4 GHz, 2.7 GHz, 5.0 GHz, 8.44 GHz.

We optimized the fitting function of the experimental number
counts distribution. The differential number counts ($dN/dS$) at
the various frequencies are well described by a multi power law
empirical distribution $Q(S)$ (see Equation \ref{Fitequation}).
The amplitudes ($A_1$ and $B_1$) are frequency dependent
parameters of the fit and have different values at each frequency.
Conversely the power law indices ($\varepsilon_1$, $\beta_1$,
$\varepsilon_2$ and $\beta_2$) have a common value at all the
frequencies.

The contribution of the UERS to the sky brightness was
evaluated by integrating the function $S(dN/dS)$ from the largest
flux ($S_{max} = 10^{2}$) of the measured sources down to the
lowest fluxes ($S_{min} = 10^{-12}$) corresponding to the expected
faintest sources. We got the brightness temperature with a
relative error bar of $\delta T_{UERS}/T_{UERS} \simeq 6-7\%$ at
$\nu =$ 151, 408, 610 and 1400 MHz, $\delta T_{UERS}/T_{UERS}
\simeq 9\%$ at $\nu =$ 5000 MHz, $\delta T_{UERS}/T_{UERS} \simeq
13\%$ at $\nu =$ 178 MHz and $\delta T_{UERS}/T_{UERS} \simeq
25-30\%$ at $\nu =$ 2700 and 8440 MHz.

We finally evaluated the spectral dependence of the point source
integrated brightness. As expected this dependence can be
described using a power law with a spectral index $\gamma_0 \simeq
-2.7$, in agreement with the frequency dependence of the flux
emitted by the {\it steep-spectrum} sources. We have also tested
the contribution of the {\it flat-spectrum} sources, adding a
second component with the slope $\gamma_1 = -2.0$. The
contribution of these sources starts to be relevant only at
frequencies above several GHz. In fact we estimated a contribution
by {\it flat-spectrum} sources $\sim 2 \%$ at 610 MHz and $\sim 9
\%$ at 5 GHz.

The above results were used to evaluate the CMB temperature
at frequencies close to 1 GHz from absolute measurements of the
sky temperature made by our group (see \cite[]{TRIS-I},
\cite[]{TRIS-II}, \cite[]{TRIS-III}).

\acknowledgments  {\bf Acknowledgements}: This work is part of the
TRIS activity, which has been supported by MIUR (Italian Ministry
of University and Research), CNR (Italian National Council of
Research) and the Universities of Milano and of Milano-Bicocca.
The authors acknowledge A. Franceschini for useful discussions and
the anonymous referee for helpfull comments to the draft.

\vfill \eject

\clearpage

\clearpage

\clearpage

\clearpage

\clearpage

\clearpage

\input{tab1.tex}

\input{tab2.tex}

\input{tab3.tex}

\input{tab4.tex}

\input{tab5.tex}

\end{document}

%% file: tab1.tex
\begin{deluxetable}{lll}
\tablecolumns{3} \tablewidth{0pc} \tablecaption{Reference papers for the extragalactic radio source counts. \label{tab1}} \tablehead{
\colhead{Frequency} & \colhead{} & \colhead{Reference} } \startdata 151 MHz & &
\cite{Baldwin_85}, \cite{Hales_88}, \cite{McGil_90}. \\
\cline{1-3}
178 MHz & & \cite{Gower_66}, \cite{Longair_66}. \\
\cline{1-3}
408 MHz & & \cite{Condon_84a}, \cite{Colla_73}, \cite{Mills_73}, \cite{Pearson_78}, \\
 & & \cite{Wall_80}, \cite{Birkinshaw_86}, \cite{Grueff_88}, \cite{Vallee_89}. \\
\cline{1-3}
610 MHz & & \cite{Condon_84a}, \cite{Wilson_82}, \cite{Katgert_85}, \\
 & & \cite{Katgert_79}, \cite{Bondi_07}. \\
\cline{1-3}
1.4 GHz & & \cite{Franceschini_89}, \cite{Katgert_73}, \cite{Katgert_74}, \\
 & & \cite{Katgert_76}, \cite{Fomalont_74}, \cite{Machalski_78}, \cite{Bondi_03}, \\
 & & \cite{Condon_82a}, \cite{Condon_82b}, \cite{Coleman_85}, \\
 & & \cite{Condon_84b}, \cite{Windhorst_85}, \cite{White_97}, \\
 & & \cite{Richards_00}, \cite{Prandoni_01}, \cite{Hopkins_03}. \\
\cline{1-3}
2.7 GHz & & \cite{Peacock_81b}, \cite{Wall_81}, \cite{Wall_85}, \\
 & & \cite{Gruppioni_97}. \\
\cline{1-3}
5 GHz & & \cite{Franceschini_89}, \cite{Partridge_86}, \cite{Pauliny_80}, \\
 & & \cite{Owen_83}, \cite{Davis_71}, \cite{Bennett_83}, \cite{Ledden_80}, \\
 & & \cite{Fomalont_84}, \cite{Pauliny_78}, \cite{Fomalont_88}, \\
 & & \cite{Prandoni_06}, \cite{Ciliegi_03}. \\
\cline{1-3}
8.44 GHz & & \cite{Toffolatti_98}, \cite{Windhorst_93}, \cite{Aizu_87}, \\
 & & \cite{Seielstad_83}, \cite{Fomalont_02}. \\
\enddata
\end{deluxetable}

%% file: tab2.tex
\begin{deluxetable}{ccccccccc}
\tablecolumns{9} \tablewidth{0pc} \tablecaption{Estimated values
of the parameters of the fit function $Q_1(S)$: $\varepsilon_1$,
$\beta_1$, $A_1$ and $B_1$. \label{tab2}} \tablehead{
\colhead{Frequency} & \colhead{} & \multicolumn{3}{c}{Parameter} &
\colhead{} &
\multicolumn{3}{c}{Parameter} \\
\cline{3-5} \cline{7-9}\\
\colhead{(MHz)} & \colhead{} & \colhead{$\varepsilon_1$} & \colhead{} & \colhead{$\beta_1$} & \colhead{} & \colhead{$A_1(\times 10^{-4})$} &
\colhead{} & \colhead{$B_1(\times 10^{-4})$} } \startdata
151 & & $-0.95 \pm 0.05$ & & $0.37 \pm 0.06$ & & $1.65 \pm 0.02 $ & & $1.14 \pm 0.04 $ \\
178 & & $\ldots$ & & $0.31 \pm 0.04^a$ & & $1.9 \pm 0.3$ & & $1.75 \pm 0.07$ \\
408 & & $-0.84 \pm 0.03$ & & $0.38 \pm 0.04$ & & $2.60 \pm 0.05 $ & & $4.6 \pm 0.1$ \\
610 & & $-0.91 \pm 0.07$ & & $0.4 \pm 0.1$ & & $3.04 \pm 0.08 $ & & $7.9 \pm 0.6$ \\
1400 & & $-0.854 \pm 0.007$ & & $0.37 \pm 0.02$ & & $4.65 \pm 0.02$ & & $23.2 \pm 0.8$ \\
2700 & & $-0.76 \pm 0.04$ & & $0.37 \pm 0.06$ & & $6 - 12 $ & & $60 \pm 2 $ \\
5000 & & $-0.83 \pm 0.05$ & & $0.29 \pm 0.06$ & & $8.4 \pm 0.4 $ & & $105 \pm 4$ \\
8440 & & $-0.78 \pm 0.05$ & & $0.47 \pm 0.15$ & & $16 - 34 $ & & $138 \pm 9$ \\
\cline{1-9} \colhead{W.Av.} & & $-0.854 \pm 0.007$ & & $0.37 \pm 0.02$ & & $\ldots$ & & $\ldots$ \\
\enddata
\tablenotetext{a}{The value of $\beta_1$ at 178 MHz has been
evaluated using only few data and for this reason it is not
included in the calculation of the average value.}
\end{deluxetable}

%% file: tab3.tex
\begin{deluxetable}{ccccccccc}
\tablecolumns{9} \tablewidth{0pc} \tablecaption{Estimated values
of the parameters of the fit function $Q_2(S)$: $\varepsilon_2$,
$\beta_2$, $A_2/A_1$ and $B_2/B_1$. \label{tab3}} \tablehead{
\colhead{Frequency} & \colhead{} & \multicolumn{3}{c}{Parameter} &
\colhead{} &
\multicolumn{3}{c}{Parameter} \\
\cline{3-5} \cline{7-9}\\
\colhead{(MHz)} & \colhead{} & \colhead{$\varepsilon_2$} &
\colhead{} & \colhead{$\beta_2$} & \colhead{} &
\colhead{$A_2/A_1$} & \colhead{} & \colhead{$B_2/B_1(\times
10^{7})$} } \startdata
 151 & & $\ldots$ & & $\ldots$ & & $0.24 \pm 0.04^a$ & & $\ldots$ \\
 178 & & $\ldots$ & & $\ldots$ & & $0.24 \pm 0.04^a$ & & $\ldots$ \\
 408 & & $\ldots$ & & $\ldots$ & & $0.24 \pm 0.04^a$ & & $\ldots$ \\
 610 & & $\ldots$ & & $\ldots$ & & $0.24 \pm 0.04^a$ & & $1.14\pm 0.4$ \\
1400 & & $-0.862\pm 0.032^a$ & & $1.47\pm 0.15$ & & $0.24 \pm 0.02$ & & $1.9\pm 0.2$ \\
2700 & & $\ldots$ & & $\ldots$ & & $ 0.24 - 0.30 $ & & $\ldots$ \\
5000 & & $-0.852\pm 0.027^a$ & & $\ldots$ & & $0.30\pm 0.04$ & & $2.3\pm 1$ \\
8440 & & $\ldots$ & & $\ldots$ & & $0.31 \pm 0.04^b$ & & $\ldots$ \\
\cline{1-9} \colhead{W.Av.} & & $-0.856 \pm 0.021$ & & $1.47\pm
0.15$ & & $\ldots$ & & $1.8\pm 0.2$ \\
\enddata
\tablenotetext{a} {Parameter estimated by using the model by
\cite{Franceschini_89}.} \tablenotetext{b} {Parameter estimated by
using the model by \cite{Toffolatti_98}.}
\end{deluxetable}

%% file: tab4.tex
\begin{deluxetable}{lllll}
\tablecolumns{5} \tablewidth{0pc} \tablecaption{Brightness
temperature of the unresolved extragalactic radio sources.
\label{tab4}} \tablehead{ \colhead{Frequency (MHz)} & \colhead{} &
\colhead{$T_{uers}$ (mK)} & \colhead{} & \colhead{$\Delta
T_{uers}$ (mK)} } \startdata
\ \ \ 151  & & \ \ \ \ 38600 & & \ \ \ \ 2300 \\
\ \ \ 178  & & \ \ \ \ 22800 & & \ \ \ \ 3000 \\
\ \ \ 408  & & \ \ \ \ 2600  & & \ \ \ \ 200  \\
\ \ \ 610  & & \ \ \ \ 910   & & \ \ \ \ 70   \\
\ \ \ 1400 & & \ \ \ \  95   & & \ \ \ \ 7    \\
\ \ \ 2700 & & \ \ \ \  11   & & \ \ \ \ $^{+4}_{-2}$ \\
\ \ \ 5000 & & \ \ \ \  3.3  & & \ \ \ \ 0.3  \\
\ \ \ 8440 & & \ \ \ \ 0.41  & & \ \ \ \ $^{+0.15}_{-0.06}$ \\
\enddata
\end{deluxetable}

%% file: tab5.tex
\begin{deluxetable}{lcccccc}
\tablecolumns{7} \tablewidth{0pc} \tablecaption{Frequency
dependence of the brightness temperature of the unresolved
extragalactic radio sources (see the text). \label{tab5}}
\tablehead{ \colhead{} & \colhead{} & \colhead{FIT1$^a$} &
\colhead{} & \colhead{FIT2$^b$} & \colhead{} & \colhead{FIT3$^c$}
} \startdata
$\nu_0$ (MHz) & & 610 & & 610 & & 610 \\
$T_0$ (mK) & & $880 \pm 28$ & & $903 \pm 30$ & & $876 \pm 22$ \\
$\gamma_0$ & & $-2.707 \pm 0.027$ & & $-2.681 \pm 0.029$ & & $-2.70$ \\
$T_1$ (mK) & & $\ldots$ & & $\ldots$ & & $18.9 \pm 0.2$ \\
$\gamma_1$ & & $\ldots$ & & $\ldots$ & & $-2.00$ \\
$\chi^2_R$ & & $1.26$ & & $0.11$ & & $0.05$ \\
$Q$ & & $0.026$ & & $0.0095$ & & $0.0031$ \\
\enddata
\tablenotetext{a}{\textit{FIT1} has been obtained using all the
experimental data and fitting a single power law.}
\tablenotetext{b}{\textit{FIT2} has been obtained excluding the
experimental data at $\nu=178$ MHz, 2.7 GHz, 8.44 GHz and fitting
a single power law.} \tablenotetext{c}{\textit{FIT3} has been
obtained excluding the experimental data at $\nu=178$ MHz, 2.7
GHz, 8.44 GHz and fitting {\it steep-spectrum} plus {\it
flat-spectrum} sources.}
\end{deluxetable}